\documentclass[aps,preprintnumbers,superscriptaddress,showpacs]{revtex4}
\usepackage{epsfig}
\usepackage{psfrag}
\usepackage{amsfonts}
\usepackage{graphicx}
\usepackage{dcolumn}
\usepackage{bm}

\begin{document}

\title{Drag force in a D-instanton background}

\author{Zi-qiang Zhang}
\email{zhangzq@cug.edu.cn} \affiliation{School of Mathematics and
Physics, China University of Geosciences(Wuhan), Wuhan 430074,
China}

\author{Zhong-jie Luo}
\email{luozhj@cug.edu.cn} \affiliation{School of Mathematics and
Physics, China University of Geosciences(Wuhan), Wuhan 430074,
China}

\author{De-fu Hou}
\email{houdf@mail.ccnu.edu.cn} \affiliation{Key Laboratory of
Quark and Lepton Physics (MOE), Central China Normal University,
Wuhan 430079, China}

%%%%%%%%%%%%%%%%%%%%%%%%%%%%%%%%%%%%%%%%
\begin{abstract}
We study the drag force and diffusion coefficient with respect to
a moving heavy quark in a D-instanton background, which
corresponds to the Yang-Mills theory in the deconfining,
high-temperature phase. It is shown that the presence of the
D-instanton density tends to increase the drag force and decrease
the diffusion coefficient, reverse to the effects of the velocity
and the temperature. Moreover, the inclusion of the D-instanton
density makes the medium less viscous.

\end{abstract}

\maketitle
%%%%%%%%%%%%%%%%%%%%%%%%%%%%%%%%%%%%%%%%
\section{Introduction}
Heavy ion collisions at Relativistic Heavy Ion Collider (RHIC) and
Large Hadron Collider (LHC) are believed to produce a new state of
matter so-called strongly-coupled quark gluon plasma (QGP). It was
shown that the life-time of QGP is very short ($\sim$5-10 fm/c),
hence direct detection of QGP is not possible. Thus, one needs to
rely on indirect measurement using suitable probes. Heavy quarks
are considered good probes to study the properties of QGP due to
their large mass and other unique properties, and there are
extensive experimental and theoretical efforts in the field of
heavy-flavor probes, for recent reviews on this topic, see e.g.
\cite{GR,PR,MD}.

Anti-de-Sitter space/confromal field theory (AdS/CFT), which maps
a $d$ dimensional quantum field theory to its dual gravitational
theory, living in $d+1$ dimensional, has yielded many important
insights for studying different aspects of QGP
\cite{Maldacena:1997re,Gubser:1998bc,MadalcenaReview}. In this
approach, the drag force on a moving heavy quark in
$\mathcal{N}=4$ supersymmetric Yang-Mills (SYM) plasma was first
studied in \cite{CP,GB}. Therein, the energy loss of the quark is
understood as the momentum flow along the string into the horizon.
Subsequently, there are many attempts to address the drag force in
this direction. For instance, the effect of chemical potential on
the drag force is discussed in \cite{ECA,LC}. The effect of
non-commutativity on the drag force is addressed in \cite{TMA}.
The finite coupling corrections corrections on the drag force are
analyzed in \cite{KBF}. The drag force in three charges
non-extremal black hole model is studied in \cite{JSA}. The drag
force in AdS/QCD models is investigated in \cite{ENA,PE,UGU}.
Other related results can be found, for example, in
\cite{DGI,SCH,MCH,ANA,KLP,SRO,SSG1}.

In fact, there is another check of gauge/gravity duality, the
correspondence between non-perturbative objects such as
instantons. It was shown \cite{ND,OA} that the Yang-Mills
instantons are identified with the D-instantons of type IIB string
theory. The near horizon limit of D-instantons homogeneously
distributed over D3-brane at zero temperature has been studied in
\cite{LHH}. The holographic dual of uniformly distributed
D-instantons over D3-brane at finite temperature has been analyzed
in \cite{BG}. It was argued that the features of D3-D(-1)
configuration are similar to QCD at finite temperature. For
instance, the chiral symmetry breaking exists in the D-instanton
background. The dual gauge theory of the background has a
confinement property with the linear quark-antiquark potential.
Thus, one expects that the results obtained from these theories
could provide qualitative insights into analogous questions in
QCD. For that reason, many quantities have been studied in the
D-instanton background, such as phase transitions \cite{BG}, light
flavor quark \cite{KG}, jet quenching parameter and heavy quark
potential \cite{ZQ1}.

In this paper, we study the drag force and diffusion coefficient
with respect to a  moving heavy quark in the D-instanton
background. More specifically, we would like to see how the
D-instanton density affects the drag force as well as the
diffusion coefficient. This is the purpose of the present work.

The organization of this paper is as follows. In the next section,
we briefly review the geometry of the D-instanton background at
finite temperature. In section 3, we study the effect of the
D-instanton density on the drag force. In section 4, we discuss
the relaxation time and the diffusion coefficient in this
background as well. In the last section, we end up with some
discussions.

%%%%%%%%%%%%%%%%%%%%%%%%%%%%%%%%%%%%%%%%

\section{Background geometry}
In this section we briefly review the D-instanton background. The
geometry is the one which is a finite temperature extension of
D3/D-instanton background given in \cite{KG}. The background has
an axion field and a five-form field strength which couples to the
D-instanton and D3, respectively. In Einstein frame the ten
dimensional super-gravity action is found to be \cite{GW,AK}
\begin{equation}
S=\frac{1}{16\pi G_{10}}\int
d^{10}x\sqrt{g}(\mathcal{R}-\frac{1}{2}(\partial\Phi)^2+\frac{1}{2}e^{2\Phi}(\partial\chi)^2-\frac{1}{6}F^2_{(5)}),\label{action}
\end{equation}
where $G_{10}$ is the 10-dimensional gravitational constant.
$\mathcal{R}$ denotes the Ricci scalar. $\Phi$ represents the
dilaton. $\chi$ refers to the axion. $F_{(5)}$ stands for the
field strength associated with Abelian gauge connection.

If one sets $\chi=-e^{-\Phi}+\chi_0$ in (\ref{action}), the
dilaton term and the axion term can cancel. After that, the
solution of (\ref{action}) can be written as \cite{KG1}
\begin{equation}
ds^2=e^{\frac{\Phi}{2}}[-\frac{r^2}{R^2}f(r)dt^2+\frac{r^2}{R^2}d\vec{x}^2+\frac{1}{f(r)}\frac{R^2}{r^2}dr^2+R^3d\Omega_5^2],\label{metric}
\end{equation}
with
\begin{equation}
e^\Phi=1+\frac{q}{r^4_t}log\frac{1}{f(r)},\qquad
f(r)=1-\frac{r_t^4}{r^4}\label{efai},\qquad
\end{equation}
where $R$ is the AdS radius.
$\lambda=g_{YM}^2N_c=\frac{R^4}{{\alpha^\prime}^2}$ with $\lambda$
the 't Hooft coupling and $\alpha^\prime$ the reciprocal of the
string tension. $\vec{x}=x_1,x_2,x_3$ are the boundary
coordinates. $r$ denotes the radial coordinate. The event horizon
is located at $r=r_t$. The boundary is $r=\infty$. The parameter
$q$ represents the D-instanton density as well as the vacuum
expectation value of the gauge field condensate. Moreover, the
temperature of the black hole is
\begin{equation}
T=\frac{r_t}{\pi R^2}.
\end{equation}

Also, note that for $q=0$ in (\ref{metric}), the
$AdS_5$-Schwarzschild metric is recovered.

\section{drag force}
In this section, we study the behavior of the drag force for the
background metric (\ref{metric}). It is known that when a heavy
quark moves in the plasma, it feels a drag force and consequently
loses energy. On the other hand, the energy loss can be described
in a dual trailing string picture \cite{CP,GB}: a heavy quark
moving on the boundary, but with a string tail into the AdS bulk.
According to this scenario, the dissipation of the heavy quark can
be depicted by the drag force, which is conjectured to be related
to a string tail in the fifth dimension.

The drag force is associated with the damping rate $\mu$ (or
friction coefficient), defined by Langevin equation,
\begin{equation}
\frac{dp}{dt}=-\mu p+f_1,
\end{equation}
subject to a driving force $f_1$. And for $dp/dt=0$, the driving
force is equivalent to a drag force $f$.

Now we discuss a heavy quark moving in one direction, e.g., $x_1$
direction. The coordinates are parameterized as
\begin{equation}
t=\tau, \qquad x_1=vt+\xi(r),\qquad x_2=0,\qquad x_3=0,\qquad
r=\sigma.\label{par}
\end{equation}

The string dynamic is captured by the Nambu-Goto action,
\begin{equation}
S=-\frac{1}{2\pi\alpha^\prime}\int d\tau d\sigma\sqrt{-g},
\label{S}
\end{equation}
where $g$ is the determinant of the induced metric with
\begin{equation}
g_{\alpha\beta}=g_{\mu\nu}\frac{\partial
X^\mu}{\partial\sigma^\alpha} \frac{\partial
X^\nu}{\partial\sigma^\beta},
\end{equation}
where $g_{\mu\nu}$ and $X^\mu$ are the brane metric and the target
space coordinates, respectively.

Substituting (\ref{par}) into (\ref{metric}), the induced metric
reads
\begin{equation} g_{tt}=-e^{\frac{\Phi}{2}}\frac{r^2f(r)}{R^2}, \qquad
g_{xx}=e^{\frac{\Phi}{2}}\frac{r^2}{R^2},\qquad
g_{rr}=e^{\frac{\Phi}{2}}\frac{R^2}{r^2f(r)},
\end{equation}
given this, the Lagrangian density is found to be
\begin{equation}
\mathcal
L=\sqrt{-g_{rr}g_{tt}-g_{rr}g_{xx}v^2-g_{xx}g_{tt}{\xi^\prime}^2}=\sqrt{e^\Phi[1-\frac{v^2}{f(r)}+\frac{r^4f(r)}{R^4}{\xi^\prime}^2]},
\end{equation}
with $\xi^\prime=d\xi/d\sigma$. As the action does not depend on
$\xi$ explicitly, the momentum is a constant,
\begin{equation}
\Pi_\xi=\frac{\partial\mathcal L }{\partial
\xi^\prime}=\xi^\prime\frac{e^{\frac{\Phi}{2}}r^4f(r)/R^4}{\sqrt{1-\frac{v^2}{f(r)}+\frac{r^4f(r)}{R^4}{\xi^\prime}^2}}=constant\label{lag},
\end{equation}
which leads to
\begin{equation}
{\xi^\prime}^2=\frac{{\Pi_\xi}^2[1-\frac{v^2}{f(r)}]}{\frac{r^4f(r)}{R^4}[e^\Phi\frac{r^4f(r)}{R^4}-{\Pi_\xi}^2]},\label{xi}
\end{equation}
note that in the right hand side of the above equation, the
denominator and numerator are both positive for large $r$ and
negative for small $r$ near the horizon. On the other hand,
${\xi^\prime}^2$ should be everywhere positive. With these
conditions, one finds that the denominator and numerator change
sigh at the same point. For the numerator, the critical point
$r_c$ satisfies
\begin{equation}
f(r_c)=v^2,
\end{equation}
which leads to
\begin{equation}
r_c=\frac{r_t}{(1-v^2)^{1/4}}.\label{rc}
\end{equation}
For the denominator, it also changes sigh at $r_c$, which yields
\begin{equation}
{\Pi_\xi}=\sqrt{e^{\Phi(r_c)}}\frac{r_c^2\sqrt{f(r_c)}}{R^2}=\sqrt{1+\frac{q}{r^4_t}log\frac{1}{v^2}}\frac{v}{\sqrt{1-v^2}}\frac{r_t^2}{R^2}.
\end{equation}

On the other hand, the current density for momentum $p_1$ is
\begin{equation}
\pi_x^r=-\frac{1}{2\pi\alpha^\prime}\xi^\prime\frac{g_{tt}g_{xx}}{-g},
\end{equation}
and the drag force is
\begin{equation}
f=\frac{dp_1}{dt}=\sqrt{-g}\pi_x^r,
\end{equation}
results in
\begin{equation}
f=-\frac{1}{2\pi\alpha^\prime}\sqrt{1+\frac{q}{r^4_t}log\frac{1}{v^2}}\frac{v}{\sqrt{1-v^2}}\frac{r_t^2}{R^2},
\end{equation}
where the minus sign implies that the drag force is against the
movement.

By using the relations
\begin{equation}
\lambda=g_{YM}^2N_c=\frac{R^4}{{\alpha^\prime}^2},\qquad
T=r_t/(\pi R^2),
\end{equation}
the drag force in the D-instanton background is obtained as
\begin{equation}
f=-\sqrt{1+\frac{q}{\pi^4R^8T^4}log\frac{1}{v^2}}\frac{\pi
T^2\sqrt{\lambda}}{2}\frac{v}{\sqrt{1-v^2}}.\label{drag}
\end{equation}

Also, the rate of energy loss in D-instanton background is found
to be
\begin{equation}
\frac{dE}{dt}=\vec{f}\cdot\vec{v}=-\sqrt{1+\frac{q}{\pi^4R^8T^4}log\frac{1}{v^2}}\frac{\pi
T^2\sqrt{g_{YM}^2N}}{2}\frac{v^2}{\sqrt{1-v^2}}.
\end{equation}

Let us discuss results. First, for $q=0$ in (\ref{drag}), the drag
force of $\mathcal{N}=4$ SYM theory \cite{CP,GB} can be
reproduced, as expected. Furthermore, to study the effect of
D-instanton density on the drag force, we plot $f/f_{SYM}$ for
various temperatures and velocities in fig.1. The left is plotted
for a low temperature ($r_t=1$) while the right is for higher one
($r_t=2$). From the figures, one can see that the drag force in
D-instanton background is larger that of $\mathcal{N}=4$ SYM
theory. Also, increasing $q$ leads to increasing $f$, opposite to
the effects of $v$ and $T$ (or $r_t$). Therefore, one concludes
that the D-instanton density has the effect of increasing the drag
force. Interestingly, it was shown \cite{ZQ1} that the presence of
the D-instanton density enhances the jet quenching parameter,
which means that regarding the energy loss, the effects of
D-instanton density on the drag force and jet quenching parameter
are consistent.

\begin{figure}
\centering
\includegraphics[width=8cm]{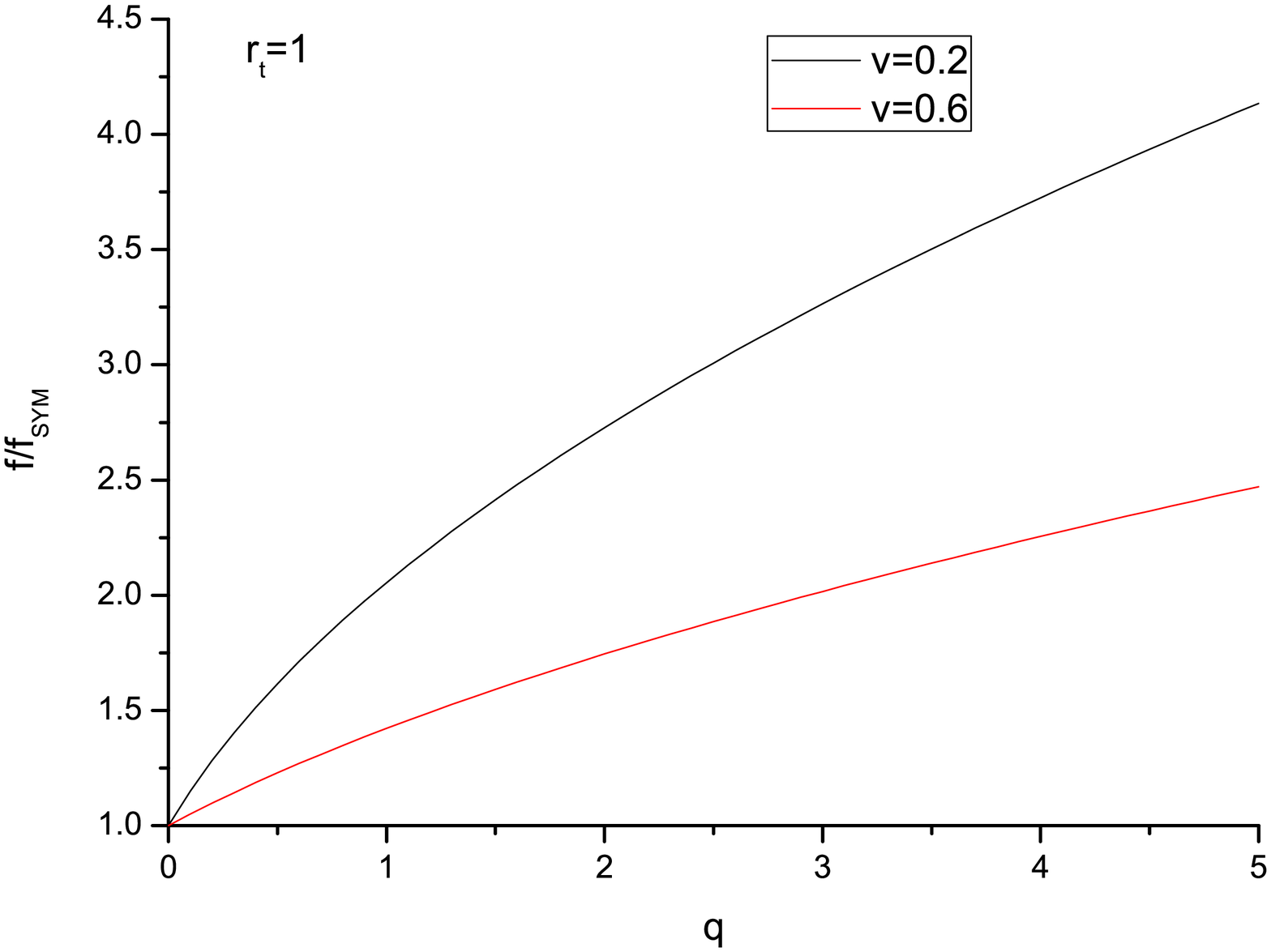}
\includegraphics[width=8cm]{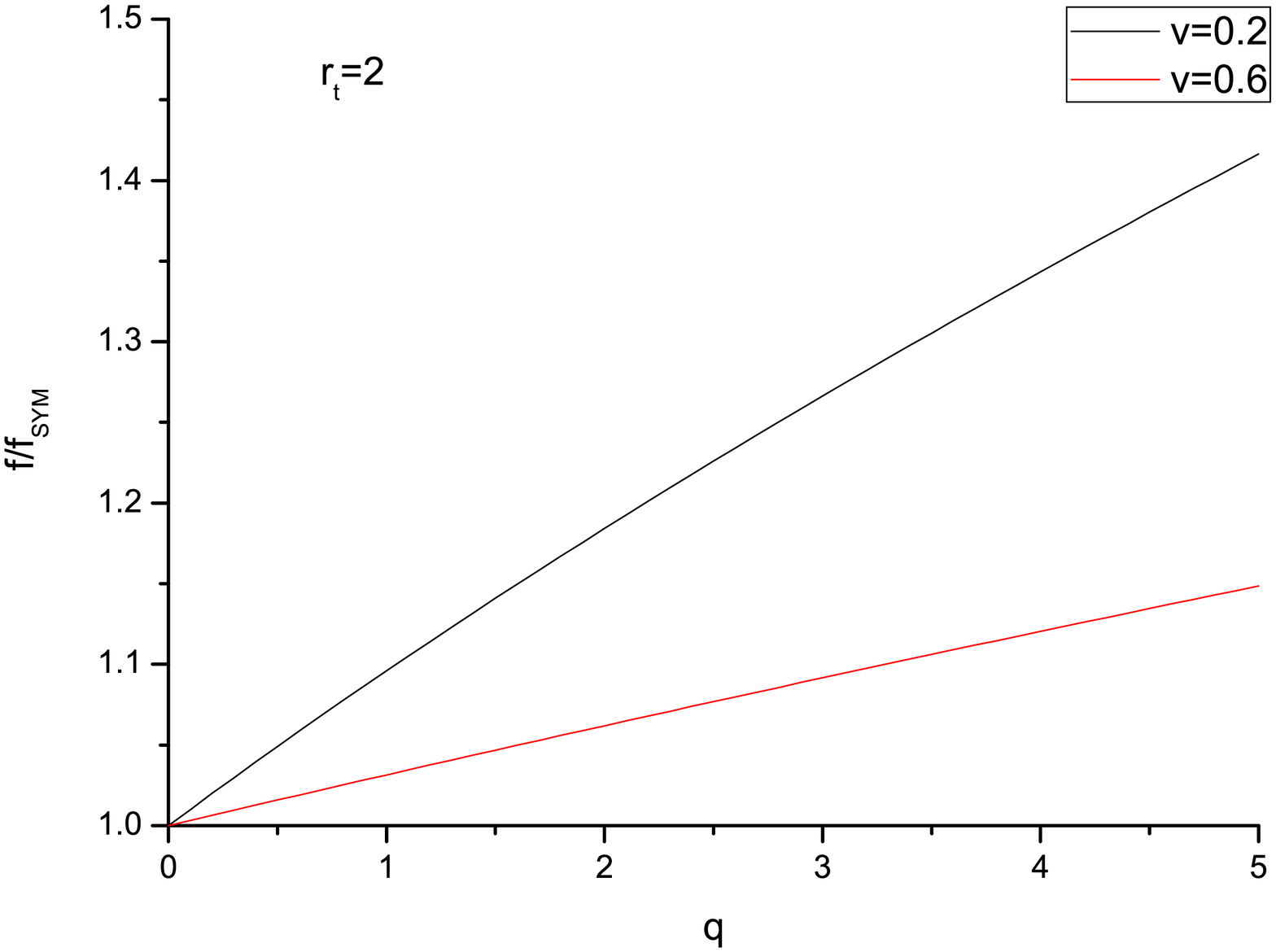}
\caption{$f/f_{SYM}$ versus q. Left: $r_t=1$. Right: $r_t=2$. In
all of the plots from top to bottom $v=0.2,0.6$, respectively.}
\end{figure}

\section{diffusion coefficient}
The diffusion coefficient, a fundamental parameter of plasma at
RHIC and LHC for heavy quarks, can be derived from the drag force.
In this section, we follow the argument in \cite{CP,GB} to study
the effect of the D-instanton density on the diffusion
coefficient.

To begin with, we recall the results of $\mathcal{N}=4$ SYM theory
in \cite{CP,GB} as follows. The drag force is
\begin{equation}
f_{SYM}=-\frac{\pi
T^2\sqrt{\lambda}}{2}\frac{v}{\sqrt{1-v^2}},\label{drag1}
\end{equation}
and the relaxation time is
\begin{equation}
t_{SYM}=\frac{2m}{\pi T^2\sqrt{\lambda}},
\end{equation}
where $m$ is the mass of the heavy quark.

The diffusion coefficient is given by
\begin{equation}
D_{SYM}=\frac{T}{m}t_{SYM}=\frac{2}{\pi T\sqrt{\lambda}}.
\end{equation}

Likewise, one can derive the relaxation time and diffusion
coefficient with the effect of the D-instanton density from
(\ref{drag}). The relaxation time is
\begin{equation}
t=\frac{1}{\sqrt{1+\frac{q}{\pi^4R^8T^4}log\frac{1}{v^2}}}\frac{2m}{\pi
T^2\sqrt{\lambda}}.
\end{equation}

The diffusion coefficient is
\begin{equation}
D=\frac{1}{\sqrt{1+\frac{q}{\pi^4R^8T^4}log\frac{1}{v^2}}}\frac{2}{\pi
T\sqrt{\lambda}}.
\end{equation}

To illustrate the effect of the D-instanton density on the
diffusion coefficient, we plot $D/D_{SYM}$ at two fixed
temperature ($r_h=1$) for two different velocities in fig.2. One
finds that the D-instanton density decreases the diffusion
coefficient, while the velocity and temperature have opposite
effect. Also, the effect of the D-instanton density on the
relaxation time is similar to that on the diffusion coefficient.

\begin{figure}
\centering
\includegraphics[width=8cm]{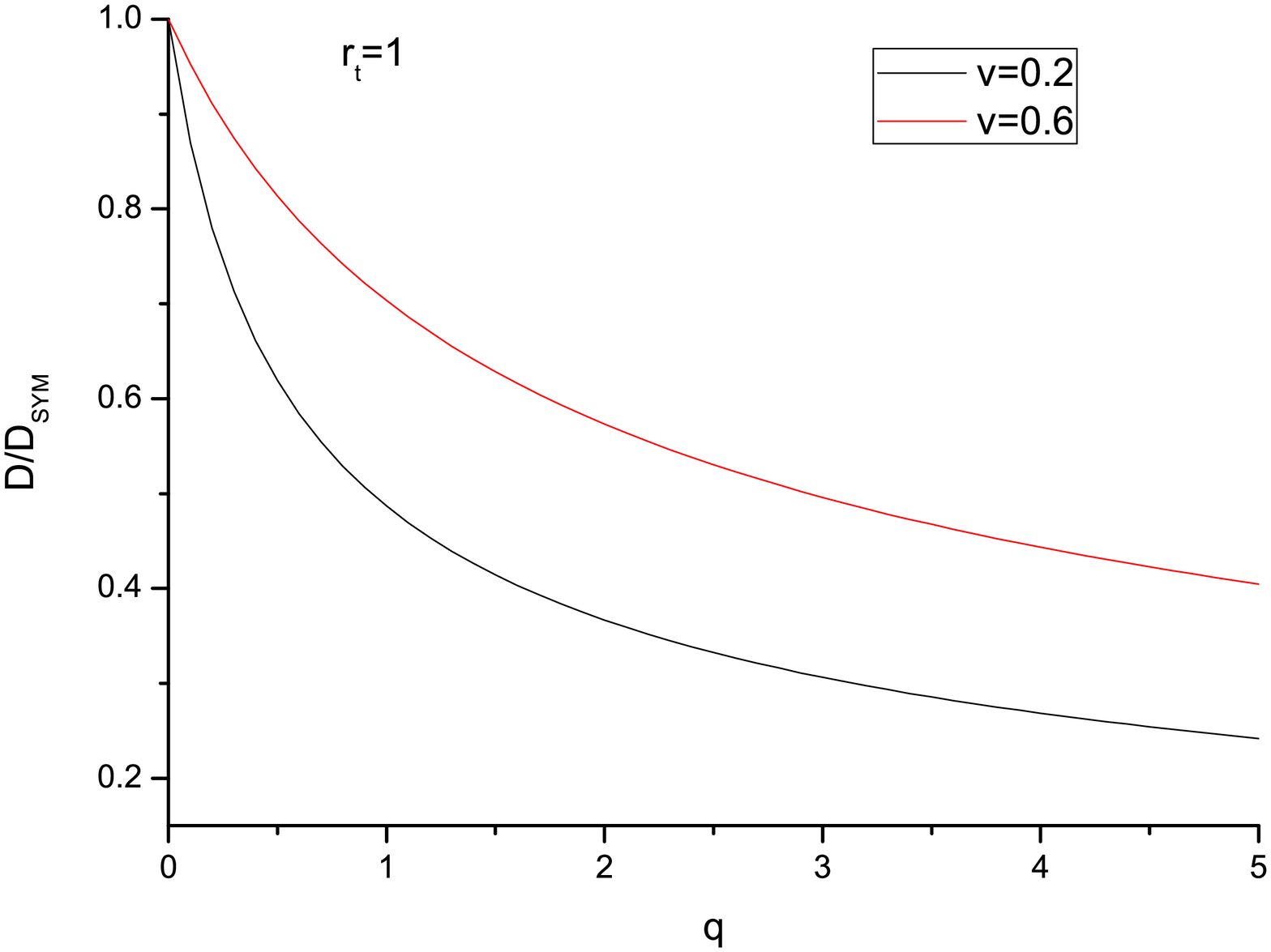}
\includegraphics[width=8cm]{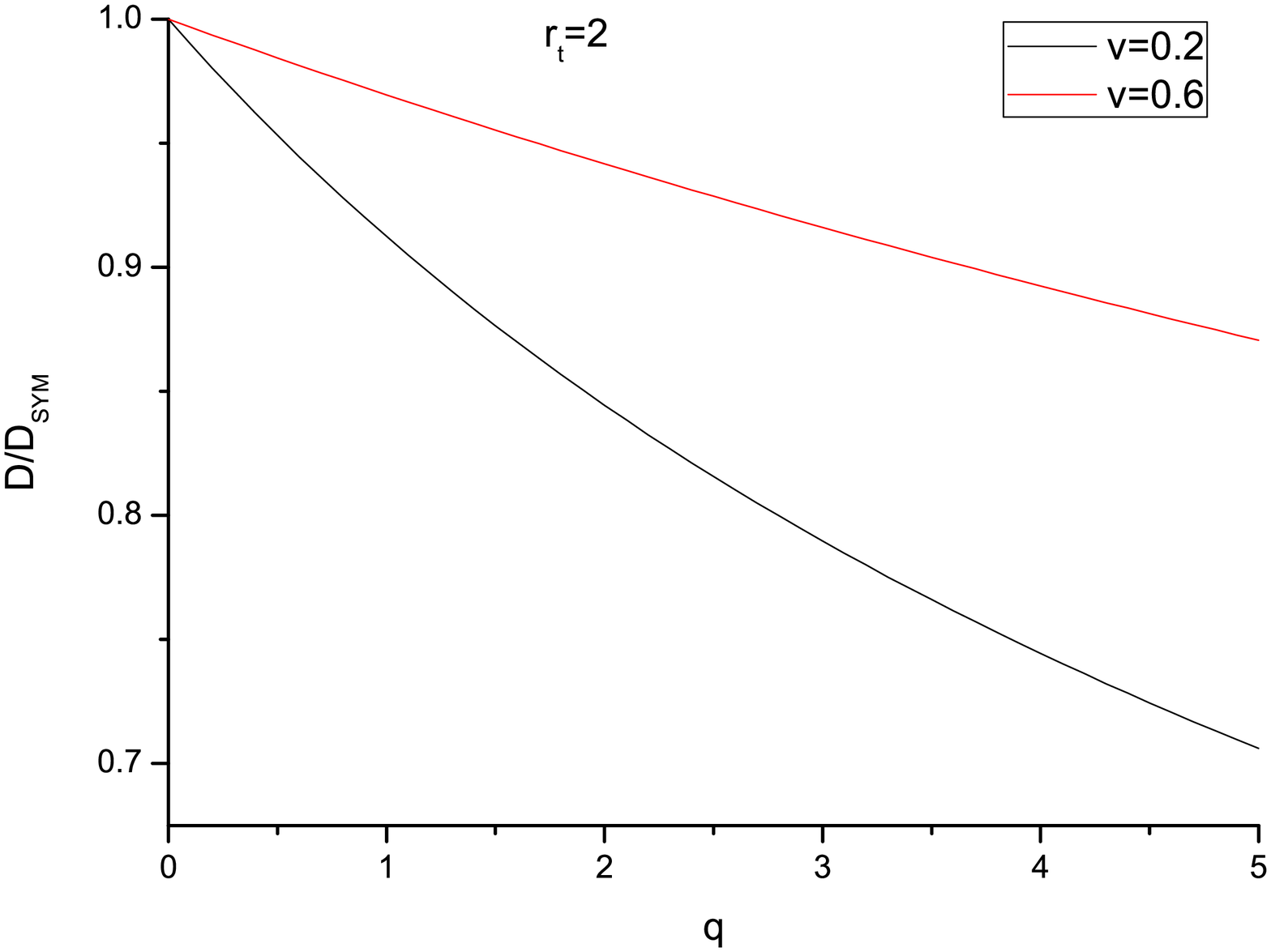}
\caption{$D/D_{SYM}$ versus q. Left: $r_t=1$. Right: $r_t=2$. In
all of the plots from top to bottom $v=0.6,0.2$, respectively.}
\end{figure}

On the other hand, one would like to discuss the mass of the heavy
quark. It was shown that \cite{JCAS} the relaxation time must be
larger than the inverse temperature
\begin{equation}
t>>\frac{1}{T}.
\end{equation}

In terms of $t=\frac{m}{T}D$, one finds
\begin{equation}
m>>\sqrt{1+\frac{q}{\pi^4R^8T^4}log\frac{1}{v^2}}\frac{\pi
T\sqrt{\lambda}}{2},
\end{equation}
which implies that the presence of the D-instanton density
increases the lower limit of the mass of the heavy quark.

\section{Conclusion}
Drag force and diffusion coefficient are two important quantities
that can be related to the energy loss in dissipative processes in
medium. In this paper, we studied these two quantities in an AdS
configuration generated by a dilaton field, corresponding to the
D-instanton density contributions. It is shown that the presence
of the D-instanton density tends to increase the drag force and
decrease the diffusion coefficient, reverse to the effects of the
velocity and the temperature. Interestingly, it was argued that
\cite{ZQ1} the D-instanton density enhances the jet quenching
parameter. Thus, one concludes that the effects of D-instanton
density on the drag force and the jet quenching parameter are
consistent.

Moreover, one can analyze the effects of D-instanton density on
the viscosity. It is known that a stronger force and associated
smaller diffusion coefficient imply a more strongly coupled
medium, closer to an ideal liquid. Since the D-instanton density
has the effect of increasing the drag force, one argues that the
presence of the D-instanton density tends to decrease the
viscosity of QGP.

\section{Acknowledgments}
This work is partly supported by the Ministry of Science and
Technology of China (MSTC) under the ¡°973¡± Project No.
2015CB856904(4). Z-q Zhang is supported by the NSFC under Grant
No. 11705166. D-f. Hou is supported by the NSFC under Grants Nos.
11735007, 11521064.

%%%%%%%%%%%%%%%%%%%%%%%%%%%%%%%%%%%%%%%%

\end{document}